\documentclass[aps,prd,twocolumn,preprintnumbers,nofootinbib,superscriptaddress,amsmath]{revtex4-2}
\usepackage[english]{babel}
\usepackage{bm}
\usepackage[utf8]{inputenc}
\usepackage[colorinlistoftodos, color=green!40, prependcaption]{todonotes}
\usepackage{amssymb}
\usepackage{graphicx}
\usepackage{natbib}
\usepackage[normalem]{ulem}

\usepackage[pdftex, pdftitle={Article}, pdfauthor={Author}]{hyperref} 

\begin{document}

\title{Cosmological Constraints on Neutrino Masses in Quintessential Inflation}

\author{Jamerson Rodrigues}
\email{jamersonrodrigues@on.br}
\affiliation{Observatório Nacional, Rio de Janeiro - RJ, 20921-400, Brasil}
\affiliation{Departamento de Física Teórica e Experimental, Universidade Federal do Rio Grande do Norte, Natal - RN, 59072-970, Brasil}
\author{Gabriel Rodrigues}
\email{gabrielrodrigues@on.br}
\affiliation{Observatório Nacional, Rio de Janeiro - RJ, 20921-400, Brasil}
\author{Felipe B. M. dos Santos}
\email{fbmsantos@on.br}
\affiliation{Observatório Nacional, Rio de Janeiro - RJ, 20921-400, Brasil}
\author{Simony Santos da Costa}
\email{simony.santosdacosta@unitn.it}
\affiliation{Dipartimento di Fisica, Università di Trento, Via Sommarive 14, 38123 Povo (TN), Italia}
\affiliation{Trento Institute for Fundamental Physics and Applications (TIFPA)-INFN, Via Sommarive 14, 38123 Povo (TN), Italia}
\affiliation{International Institute of Physics, Universidade Federal do Rio Grande do Norte, Campus Universitário, Lagoa Nova, Natal-RN 59078-970, Brasil}
\author{Jailson Alcaniz}
\email{alcaniz@on.br}
\affiliation{Observatório Nacional, Rio de Janeiro - RJ, 20921-400, Brasil}
\date{\today}

\begin{abstract}
Quintessential inflation provides a unified description of the early and late accelerated phases of the Universe, linking the inflationary epoch to the present-day dark energy–dominated era through a single scalar degree of freedom. In this work, we explore the implications of this unification for cosmological constraints on the sum of neutrino masses. Focusing on the $\alpha$-attractor scenario, we implement the model in a modified version of the Boltzmann solver CLASS to compute the relevant cosmological observables and perform a Bayesian parameter estimation analysis using data from the cosmic microwave background (CMB), baryon acoustic oscillations (BAOs), and Type Ia supernovae. The model naturally breaks the degeneracy between the dark energy equation of state and the total neutrino mass, yielding tight upper bounds of $\sum m_\nu< 0.067$ eV for flat spatial geometry and $\sum m_\nu< 0.116$ eV when curvature is included. We also provide forecasts for future probes, showing that the Simons Observatory, LiteBIRD, and Euclid configurations may reduce the uncertainty on $\sum m_\nu$ by $\approx 9\%$, while the precision on the quintessential parameter $\alpha_{QI}$ is improved by $\approx 72\%$. These results highlight the importance of consistently accounting for neutrino mass when assessing the viability of extensions to the standard cosmological model.

\end{abstract}

\maketitle

\section{Introduction}

In standard cosmology, the observable Universe originated in an early phase of accelerated expansion, dubbed Inflation, whose dynamics is commonly described by a scalar field known as the inflaton. The transition to a radiation-dominated Universe is achieved through the reheating mechanism, where the energy stored in the inflaton condensate is transferred to the standard particles. The subsequent expansion of the cosmos is dominated by an ordinary fluid composed initially of radiation and, as this fluid cools down, by a cold dark matter component. More recently, observations of distance–redshift relation of Type Ia supernovae revealed a late accelerated expansion~\cite{SupernovaSearchTeam:1998fmf,SupernovaCosmologyProject:1998vns} compatible with the description of an exotic component, generically dubbed dark energy (DE), whose energy density remains constant, or nearly so, as the Universe expands.

The scenario drawn from the cosmological model enables the use of probes from distinct moments of the cosmic evolution to constrain the very same parameters. A remarkable example is the present rate of the cosmic expansion, $H_0$.  One could choose to constrain the Hubble parameter indirectly from the angular scale of the acoustic peaks in the cosmic microwave background~\cite{Planck:2018vyg}, for example, or by the observations of the type Ia supernovae light curves and the cosmic ladder techniques~\cite{Riess:2020fzl}.  While such complementarity can be useful to break possible degeneracies between the cosmological model parameters, the increasing amount and precision of cosmological data open up room for tensions between the values inferred from different probes.  In the former example, the tension between early and late time probes of $H_0$ ranges between $4\sigma \, - 6\sigma$~\cite{DiValentino:2021izs}.

The cohesion  between the different cosmological eras can also be explored from the  model building  perspective. One particularly interesting example  can be found in the quintessential inflation (QI) models~\cite{Peebles:1998qn,Peloso:1999dm,Dimopoulos:2001ix,Tashiro:2003qp}. In this scenario, inflation and quintessence are unified in the same framework, with the Lagrangian of the inflaton/quintessence field prescribing the characteristics of the primordial and late Universe. By connecting early- and late-time cosmic expansion, this class of models enables the use of different probes to infer conceptually uncorrelated parameters, further increasing precision. Such a feature is already explored in the literature \cite{Geng:2015fla,Akrami:2017cir,Akrami:2020zxw,Dimopoulos:2017zvq,Geng:2017mic,Alestas:2024eic,Giare:2024sdl,Rodrigues:2025ezl}, where particularly in \cite{Akrami:2020zxw} an increase in the forecasted precision of the spectral index, $n_s$, was obtained.

In this context, QI models can help disentangle well-known parameter degeneracies. At small scales, the gravitational backreaction of massive neutrinos affects the clustering of cold dark matter in a way similar to dark energy, slowing down the late-time growth of matter perturbations (see \cite{Lesgourgues:2006nd} for more details). A degeneracy is then present on the effects of  $\sum m_{\nu}$ and the equation of state $w$. Phenomenological parametrizations of $w$, which have recently shown a preference for a phantom crossing with $w<-1$, lead to a loosening of the bounds on $\sum m_{\nu}$. On the other hand, when considering physical models for dark energy, where the equation of state is, by construction, constrained to never cross the phantom divide, so $w \geq -1$, the bounds on $\sum m_{\nu}$ are tightened~\cite{Hannestad:2005gj,RoyChoudhury:2019hls,Lorenz:2017fgo,Du:2024pai,Rodrigues:2025hso}. A representative example of this pattern can be found by comparing the results obtained in ~\cite{Du:2024pai} and \cite{Rodrigues:2025hso}. The first obtained $\sum m_\nu < 0.147$ eV for the analysis of the CPL parameterization of dark energy in light of CMB, BAO, and supernovae data, while the second obtained a much more restrictive result, $\sum m_\nu < 0.0709$ eV, following the analysis of a general parameterization of thawing quintessence models with the same dataset.

From a particle physics perspective, the oscillatory behavior of the three neutrino flavors is observed in solar and atmospheric neutrino measurements. This points to the existence of neutrino mass terms in the neutrino Lagrangian, despite predictions from the Standard Model of particles. Oscillation experiments are insensitive to the absolute mass scale of the neutrinos, but are useful for probing their squared mass differences. For atmospheric neutrinos, one can settle $\Delta m_{32}^2 \approx 2.45 \times 10^{-3}\; \text{eV}^2$, assuming the Normal Ordering (NO), or $\Delta m_{32}^2 \approx -2.53 \times 10^{-3}\; \text{eV}^2$ for the Inverted Ordering (IO), while from solar neutrinos one obtains $\Delta m_{21}^2 \approx 7.5 \times 10^{-5}\; \text{eV}^2$~\cite{ParticleDataGroup:2024cfk}. This configuration enables at least two non-relativistic species of neutrinos, while lower limits on the sum of their masses can be estimated. For the NO one may have  $\sum m_\nu > 0.06\; \text{eV}$, while IO requires  $\sum m_\nu > 0.10\; \text{eV}$. 

In a scenario where cosmological upper limits for neutrino masses are increasingly approaching the lower limits obtained when considering oscillation experiments~\cite{Palanque-Delabrouille:2019iyz,DiValentino:2021hoh,DiValentino:2022njd,DESI:2024hhd,DESI:2025ejh}, the analysis of neutrino masses in the context of quintessential inflation may emerge as a crucial test for the viability of these models.

In this work, we analyze the cosmological constraints on the total neutrino masses in the framework of the $\alpha$-attractor quintessence inflation model. The $\alpha-$attractor models were initially conceived as an inflationary scenario based on supergravity theories \cite{Kallosh:2013hoa,Kallosh:2013yoa,Kallosh:2014rga}. The K\"{a}hler potential in these frameworks entails a non-canonical pole structure for the scalar field, enabling an attractor solution in the inflationary dynamics that connects the inflationary observables to the curvature of the K\"{a}hler manifold, $\alpha$.   Such features can be understood geometrically as the stretching of the potential energy in the asymptotic values of the canonical field, independently of the original form of the scalar potential. Remarkably, the same mechanism that enables the proper description of primordial inflation can also trigger the late-time acceleration, as we will explore later.

In particular, we perform a Monte-Carlo Markov Chain (MCMC) analysis of this scenario in light of current CMB, baryon acoustic oscillations, and supernovae data. The forecasted sensitivity of the model to the configurations of future large-scale structure and CMB surveys has also been considered. Specifically, we use data from Planck, DESI DR2 BAO, and Pantheon+ for the real-data analysis, while the forecasts are made for the next-generation CMB experiments, LiteBIRD and Simons Observatory, combined with the Euclid survey configurations.

This paper is organized as follows. In Section~\ref{Sec:2}, we review the main features of the $\alpha$-attractor quintessence inflation model. Section~\ref{Sec:3} describes the methodology and data used in the analysis. In Section~\ref{Sec:4}, we present the constraints obtained from both real and simulated data. Finally, Section~\ref{Sec:5} summarizes our conclusions.

\section{The $\alpha-$attractor Quintessential Inflation}\label{Sec:2}

The underlying Lagrangian density governing the inflaton dynamics has the general form,
\begin{equation}
    \mathcal{L}=\sqrt{-g}\left(\frac{M_p^2}{2}R - \frac{1}{2}\frac{\partial_\mu\phi\partial^\mu\phi}{\left(1-\frac{\phi^2}{6\alpha_{QI}}\right)^2} - V(\phi)\right),
\end{equation}
where $M_p$ is the reduced Planck mass. The typical non-canonical kinetic term has a pole structure controlled by $\alpha_{QI}$ which, in turn, is defined by the curvature of the scalar manifold in the K\"{a}hler potential \cite{Kallosh:2013yoa}. 
 
The features of the $\alpha-$attractor models are better understood in terms of the canonical field, $\varphi$, obtained according to the relation,
\begin{equation}\label{eq:7}
   \phi\equiv\sqrt{6\alpha_{QI}}\tanh\frac{\varphi}{\sqrt{6\alpha_{QI}}}.
\end{equation}
Such transformation drives the poles to large amplitudes of the canonical field, $|\varphi|\to\infty$, yielding asymptotic flat directions for the potential energy when expressed in terms of the canonical variable ~\cite{Kallosh:2013yoa,Linde:2016uec}. Those are the key ingredients that make the $\alpha-$attractor structure suitable for inflation and quintessence. For the appropriate choice of the underlying supergravity theory, the scalar potential may develop an inflationary plateau in the limit of large and positive field, $\varphi \rightarrow +\infty$, while acquiring a runaway behavior in the negative asymptotic values, $\varphi \rightarrow -\infty$, where the scalar field drives the late accelerated expansion. 

The inflationary stage is dictated, as usual, by the slow-roll parameters, $\epsilon,\ \eta$. In particular, the boundary conditions for the metric perturbations are described in terms of the power-law expansion of the primordial power spectrum. For scalar perturbations,
\begin{equation}
    \ln{\frac{P_\mathcal{R}(k)}{P_\mathcal{R}(k_\star)}} = (n_s-1) \ln\left({\frac{k}{k_\star}}\right) + \frac{\alpha_s}{2}\ln\left({\frac{k}{k_\star}}\right)^2 + \ldots , \label{Eq:06}
\end{equation}
and similarly for the tensor perturbations, where $k_\star$ is the pivot scale and the ellipse portrays the higher order terms. When the slow-roll regime is considered, $\epsilon,\, \eta \ll 1$, the spectral index and the tensor-to-scalar ratio are computed in terms of the slow-roll parameters,
\begin{equation}
    n_s = 1 - 6\epsilon + 2\eta, \quad r=16\epsilon,
\end{equation}
respectively. One especially appealing feature of this inflationary scenario follows from its attractor behavior, which allows for general predictions for the inflationary observables~\cite{Linde:2016uec},
\begin{equation}
    n_s\simeq 1-\frac{2}{N_\star}, \quad r\simeq\frac{12\alpha_{QI}}{N_\star^2},
    \label{eq:Slow_Rel}
\end{equation}
where $N_\star$ is the number of e-folds, $N \equiv\ln(a)$. In particular, the relation of $r$ to the Lagrangian parameter $\alpha_{QI}$ enables one to tune the spectral index and the tensor-to-scalar ratio simultaneously to the sweet-spot of Planck's observations \footnote{See \cite{Rodrigues:2020fle} for a MCMC analysis of the inflationary $\alpha$-attractor scenario with a Higgs-like potential.}~\cite{Planck:2018vyg}.  To complete the description of the primordial power spectrum, one may consider the amplitude of the scalar perturbations,
\begin{equation}
    P_\mathcal{R}(k_\star)\equiv A_s(k_{\star}) =\left.\frac{V}{24M^4_P\pi^2\epsilon}\right|_{\varphi=\varphi_\star},\label{Eq:A_s}
\end{equation}
where the field amplitude $\varphi_\star$ is computed at the time of horizon crossing for the pivot scale. 

Unlike the other inflationary parameters, $A_s$ is particularly sensitive to the potential energy amplitude. Here, we consider one of the simplest forms of the potential energy that can address inflation and quintessence\footnote{Near the kinetic pole, canonical normalization implies $\phi=\sqrt{6\alpha_{QI}}\big(1-2e^{-2\varphi/\sqrt{6\alpha_{QI}}}+\cdots\big)$. As a result, any regular (e.g. polynomial) $V(\phi)$ near $\phi\simeq\sqrt{6\alpha_{QI}}$ is mapped into a plateau with exponential corrections characteristic
of $\alpha$-attractor models~\cite{Kallosh:2013yoa,Linde:2016uec}.
}~\cite{Akrami:2017cir,AresteSalo:2021wgb,Dimopoulos:2017tud},
\begin{eqnarray}
    V(\varphi) =  M^2 e^{\gamma\left(\tanh\varphi/\sqrt{6\alpha_{QI}}-1\right)}. \label{Eq:Pot_Cano}
\end{eqnarray}
This potential assumes a step-like form, with the inflationary period taking place in the positive large field regime, where $V \rightarrow M^2 $. According to the Klein-Gordon equation, the field will eventually roll down the potential hill into the negative region of field space. In the limit of large negative amplitudes, it will mimic a cosmological constant, $V\rightarrow M^2 e^{-2\gamma}$, establishing the late accelerated expansion of the Universe. The steepness of the potential function is tuned by $\gamma$ in order to reproduce the correct vacuum energy density observed today, $\rho_\Lambda \sim 2.5 \times 10^{-11}$ eV$^4$ \cite{Planck:2018vyg}. Meanwhile, the measured value of $A_s$ fixes the amplitude of the scalar potential through \eqref{Eq:A_s}, giving
\begin{equation}
    \frac{M^2}{M_P^4}=\frac{144\pi^2\alpha_{QI} N_\star A_s}{(2N_\star - 3\alpha_{QI})^3}.
\end{equation}

This set of conjectures brings the early and late-time conditions for accelerated expansion into agreement. It may also help to disentangle well-known degeneracies, especially the one relating the total neutrinos mass and the state parameter of dark energy, once one considers data from the early and late-time Universe. In the following, we will detail the numerical methods performed to constrain the model and cosmological parameters.

\section{Methodology}\label{Sec:3}

In this work, we perform an MCMC analysis to obtain the posterior probability distribution of the cosmological parameters. In particular, the \texttt{Cobaya} \cite{Torrado:2020dgo} sampler code is used to build the chains for the analysis with real data, following the Metropolis-Hastings algorithm. The predictions for cosmological observables are obtained from the Cosmic Linear Anisotropy Solving System (\texttt{CLASS}) \cite{Diego_Blas_2011} code, which has been modified to solve the Boltzmann equations for the quintessential $\alpha$-attractor model. The main modifications are performed in the primordial and background modules of \texttt{CLASS}, in order to implement the potential energy in Eq. \eqref{Eq:Pot_Cano}.

We also perform an MCMC analysis with simulated data in order to forecast the sensitivity of the total neutrino mass, in the context of the quintessential inflation scenario, to future CMB and LSS experiments. In this case, the \texttt{MontePython} \cite{Brinckmann:2018cvx} code is used to sample the likelihood of the mock data. For both real and simulated data analysis, the covariance matrix is updated until the Gelman-Rubin convergence criterion reaches the values of $R-1 < \mathcal{O}(10^{-2})$. Finally, the chains obtained are analyzed and plotted using the \texttt{GetDist} code \cite{Lewis:2019xzd}. Below, we discuss the details of the cosmological model and the datasets used in this analysis.

\subsection{The Model}

The base parameters utilized in the Bayesian parameter estimation are specified as follows,
\begin{eqnarray}
   \left( \Omega_\mathrm{b} h^2 ,\, \Omega_\mathrm{c} h^2, \, 100\theta_\mathrm{s},\, \tau_\mathrm{reio},\, \ln(10^{10} A_\mathrm{s}),\, n_\mathrm{s},\right. \nonumber \\
   \alpha_\mathrm{QI},\, \sum m_\nu  ), \label{Eq:CosmI}
\end{eqnarray}
where $\Omega_\mathrm{b} h^2$ and $\Omega_\mathrm{c} h^2$ represent the weighted abundances of baryons and cold dark matter, respectively; $100\theta_s$ denotes the angular acoustic scale; $\tau_{reio}$ is the optical depth; $\ln(10^{10}A_s)$ indicates the amplitude of scalar perturbations; $n_s$ is the spectral index; $\alpha_{QI}$ refers to the $\alpha$ parameter from $\alpha$-attractors, and $\sum m_{\nu}$ corresponds to the total neutrino mass. Hereafter dubbed $\mathrm{Cosmo\, I}$, this is the minimal set of quantities capable of describing the observables of a flat Universe, including the damping effect of the neutrinos' free-streaming and the angular distances in the latter Universe. 

To consider the effects of a non-zero curvature, we also include the energy component associated with the curvature of space-time, $\Omega_K$, as a varying parameter in the MCMC analysis, 
\begin{eqnarray}
   \small  ( \Omega_\mathrm{b} h^2 ,\, \Omega_\mathrm{c} h^2\, \, 100\theta_\mathrm{s},\, \tau_\mathrm{reio},\, \ln(10^{10} A_\mathrm{s}),\, n_\mathrm{s},\nonumber \\ \Omega_K,, \alpha_\mathrm{QI},\, \sum m_\nu  ), \label{Eq:CosmII}
\end{eqnarray}
hereafter called $\mathrm{Cosmo\, II}$  model. The curvature is particularly important in the late-time Universe as it influences the cosmic expansion and, consequently, displaces the angular distance between the observer and the source of the signal. The motivation for exploring curvature in the cosmological model is related to its geometric degeneracy with the total mass of the neutrino, which may enable the loosening of the constraints on $\sum m_\nu$. Finally, we used flat prior distributions in the analysis of real and simulated data, with the intervals specified in Table \eqref{Tab_P}.

\begin{table*}
\centering
\begin{tabular}{ l  c  c}

\hline \hline

 Parameter &  Cosmo I & Cosmo II \\
\hline
{ $\Omega_\mathrm{b} h^2$}  & $\left[0.005\, ,\, 0.1\right]$ & $\left[0.005\, ,\, 0.1\right]$\\
{ $\Omega_\mathrm{c} h^2$}  & $\left[0.001\, ,\, 0.5\right]$ & $\left[0.001\, ,\, 0.5\right]$\\
{ $100\theta_\mathrm{s}$}   & $\left[0.5\, ,\, 2.0\right]$ & $\left[0.5\, ,\, 2.0\right]$\\
{ $\tau_\mathrm{reio}$}  & $\left[0.01\, ,\, 0.2\right]$ & $\left[0.01\, ,\, 0.2\right]$\\
{ $n_\mathrm{s}   $}  & $\left[0.8\, ,\, 1.1\right]$ & $\left[0.8\, ,\, 1.1\right]$\\
{ $\ln(10^{10} A_\mathrm{s})$}  & $\left[1.61\, ,\, 3.91\right]$ & $\left[1.61\, ,\, 3.91\right]$\\
$\sum m_\nu [\text{eV}]  $  & $\left[0\,\, ,\, 1.05\,\right]$ & $\left[0\,\, ,\, 1.05\,\,\right]$\\
{$\alpha_\mathrm{QI}$}  &  $\left[0.01\, ,\, 5.0\right]$  & $\left[0.01\, ,\, 5.0\right]$\\
{$\Omega_K$}  & --- & $\left[-0.3\, ,\, 0.3\right]$\\
\hline
\end{tabular}
\caption{Prior distribution intervals used in the MCMC analysis of the models $\mathrm{Cosmo\, I}$ and $\mathrm{Cosmo\, II}$.}
\label{Tab_P}
\end{table*}

In time, it is useful to remind that the quintessential $\alpha$-attractor scenario reconciles early and late expansion with the same physical mechanism. Numerically, this is accomplished by the shooting process, which adjusts one of the parameters of the quintessential potential to recover the correct budget equation at later times. In this work, the \texttt{CLASS} code was modified to perform shooting with the parameter $\gamma$ of \eqref{Eq:Pot_Cano}, making it a derived parameter in the MCMC analysis. Another important quantity in the analysis of the late-time behavior of the quintessential field is its initial position, or freezing value, $\varphi_F$ \cite{Akrami:2017cir,Akrami:2020zxw}. The possible values of $\varphi_{F}$ vary essentially due to the reheating mechanism that takes place after inflation. For our purposes, we follow \cite{Akrami:2020zxw,Rodrigues:2025ezl} and fix $\varphi_{F} = -10$ in our analysis, which corresponds to the case of instant reheating. We expect the effect of the variation in $\varphi_{F}$ to be marginal on the constraints of the total neutrino mass, since the latter is mainly related to post-recombination physics.

\subsection{The Data}

\subsubsection{Real data}

As previously discussed, we can integrate early and late-time observations to explore how QI models can help address parameter degeneracies. In this context, we have combined observations from both early and late periods. For the early times, we utilized the latest Planck data release (PR4), processed using the NPIPE pipeline, which includes measurements of temperature (TT), polarization (EE), and the cross-correlation (TE)~\cite{rosenberg22, Planck:2018nkj}. The temperature power spectrum at low multipoles is modeled using the publicly available Commander likelihood. For the CMB lensing, we use the
Planck PR4 likelihood, obtained through the reconstruction of the lensing potential maps~\cite{Carron:2022eyg}.

The late-time measurements include SNe Ia data from the Pantheon+ catalog~\cite{Brout_2022, Scolnic_2022} and the latest geometric BAO observations from the DESI DR2~\cite{DESI:2025zgx}. The BAO measurements can determine, at different redshifts, the angular scale imprinted in the large-scale structures at the baryon drag epoch, thereby providing constraints on the matter content of the Universe. It is important to note that BAO data alone cannot distinguish between the effects of massive neutrinos and those of the standard cold dark matter component. For this reason, it is necessary to combine them with CMB measurements. These can break the degeneracy between $\Omega_\nu$ and $\Omega_c$. They also allow one to constrain the total neutrino mass, $\sum m_\nu$. This is because massive neutrinos alter the time of matter–radiation equality and suppress the growth of small-scale perturbations through free-streaming effects~\cite{Lesgourgues:2006nd}.

In our analyses, we considered the CMB dataset and the combination of CMB, BAO, and SNe Ia.

\subsubsection{Simulated data}

For the forecast analysis, we consider a combined analysis of future CMB and galaxy clustering data. For the CMB part, we include the temperature ($T$) and polarization ($E$ and $B$) modes, which are jointly described by the covariance matrix of angular power spectra,
\begin{equation}
   \textbf C_\ell = \begin{pmatrix}
C_\ell^{TT} + N_\ell^{TT} & C_\ell^{TE} & 0\\
C_\ell^{TE} & C_\ell^{EE} + N_\ell^{EE} & 0\\
0 & 0 & C_\ell^{BB} + N_\ell^{BB}\\ 
\end{pmatrix},
\end{equation}
where $C_\ell$ denotes the theoretical temperature, polarization, and cross-correlation power spectra, while $N_\ell$ represents the corresponding instrumental noise. The noise term for each channel is modeled as
\begin{equation}
N_{\ell}^{XX} = \sigma_X^2
\exp\left[\frac{\ell(\ell+1)\theta_{\mathrm{FWHM}}^2}{8\ln 2}\right],
\label{eq:12}
\end{equation}
with $\sigma_X$ being the noise level for temperature or polarization, and $\theta_{\mathrm{FWHM}}$ the full width at half maximum of the beam, expressed in radians. 

For galaxy surveys, one is interested in computing the galaxy power spectrum, modeled as
\begin{equation}
\begin{split}
P_{\text{obs}}(k_{\text{fid}}, \mu_{\text{fid}} z) 
&= \frac{1}{q_\perp^2(z) q_\parallel(z)}
\Bigg\{ \frac{\left[b\sigma_8(z) + f(k, z)\sigma_8(z)\mu^2 \right]^2}{1 + f(k,z)^2 k^2 \mu^2 \sigma_p^2(z)} \\
&\quad \times \frac{P_{\text{dw}}(k, \mu; z)}{\sigma_8^2(z)} F_z(k,\mu;z)
+ P_s(z) \Bigg\},
\end{split}
\end{equation}
where $f(k,z)$ is the linear growth rate, $P_{\text{dw}}(k,\mu;z)$ the de-wiggled (BAO-damped) power spectrum, and $P_s(z)$ the shot noise.  The model incorporates three key physical effects:  
(i) the Alcock–Paczynski effect, represented by the prefactor $\frac{1}{q_\perp^2 q_\parallel}$, with $q_\perp = D_A(z)/D_{A,\text{fid}}(z)$ and $q_\parallel = H_{\text{fid}}(z)/H(z)$;  
(ii) redshift-space distortions (Kaiser effect), through the factor $[b(z)\sigma_8(z) + f(k,z)\sigma_8(z)\mu^2]$; and  
(iii) the Fingers-of-God suppression, encoded in the denominator term $1 + f(k,z)^2 k^2 \mu^2 \sigma_p^2(z)$, which accounts for small-scale random motions.  
The redshift uncertainty is modeled by the damping term $F_z(k,\mu;z) = \exp[-k^2 \mu^2 \sigma_r^2(z)]$, with $\sigma_r(z) = c\sigma_z/H(z)$.  

In what follows, we describe the specific surveys used in our forecast analysis:

\begin{table*}
\centering
\begin{tabular}{ l  c  c}

\hline \hline
 Parameter &  CMB &  CMB + DESI + Pantheon+ \\
\hline
{ $\Omega_\mathrm{b} h^2$}  & $0.02216\pm 0.00014  $ & $0.02232\pm 0.00013$\\
{ $\Omega_\mathrm{c} h^2$}  & $0.1198\pm 0.0012 $ & $0.11786\pm 0.00070$\\
{ $100\theta_\mathrm{s}$}   & $1.04172\pm 0.00024  $ & $1.04189\pm 0.00023$\\
{ $\tau_\mathrm{reio}$}  & $0.0545\pm 0.0071$ & $0.0566\pm 0.0071  $\\
{ $n_\mathrm{s}   $}  & $0.9629\pm 0.0040 $ & $0.9678\pm 0.0034 $\\
{ $\ln(10^{10} A_\mathrm{s})$}  & $3.042\pm 0.014$ & $3.042\pm 0.014$\\
$\sum m_\nu$  & $< 0.252\,\,\text{eV}$ & $< 0.0670\,\,\text{eV}$\\
{ $\alpha_\mathrm{QI}$}  & $< 3.45 $ & $1.70^{+1.0}_{-0.41}$\\

\hline
\end{tabular}
\caption{1$\sigma$ projections and $2\sigma$ upper limits of the posterior distribution of the $\mathrm{Cosmo\, I}$ model parameters.}
\label{Tab1}
\end{table*}

\begin{table*}
\centering
\begin{tabular}{ l  c  c}

\hline \hline

 Parameter &  CMB &  CMB + DESI + Pantheon+ \\
\hline
{ $\Omega_\mathrm{b} h^2$}  & $0.02219\pm 0.00015$ & $0.02221\pm 0.00014  $\\
{ $\Omega_\mathrm{c} h^2$}  & $0.1190\pm 0.0013 $ & $0.1193\pm 0.0012 $\\
{ $100\theta_\mathrm{s}$}   & $1.04179\pm 0.00025 $ & $1.04175\pm 0.00024 $\\
{ $\tau_\mathrm{reio}$}  & $0.0473^{+0.0082}_{-0.0070}$ & $0.0544\pm 0.0072 $\\
{ $n_\mathrm{s}   $}  & $0.9646\pm 0.0045 $ & $0.9642\pm 0.0042 $\\
{ $\ln(10^{10} A_\mathrm{s})$}  & $3.024^{+0.018}_{-0.015} $ & $3.041\pm 0.014  $\\
$\sum m_\nu  $  & $< 0.629\,\,\text{eV} $ & $< 0.116\,\,\text{eV} $\\
{$\alpha_\mathrm{QI}$}  &  ---  & $1.69^{+1.0}_{-0.44}$\\
{$\Omega_K$}  & $-0.022^{+0.014}_{-0.010}  $ & $0.0021\pm 0.0013          $\\
\hline
\end{tabular}
\caption{1$\sigma$ projections and $2\sigma$ upper limits of the posterior distribution of the $\mathrm{Cosmo\, II}$ model parameters.}
\label{Tab2}
\end{table*}

\begin{itemize}

\item \textbf{Simons Observatory:} The Simons Observatory \cite{SimonsObservatory:2018koc} is a ground-based survey located in Chile, whose focus is on smaller scales in the CMB spectra, expected to cover $40\leq\ell\leq5000$ in the multipole range, thus being complementary to the current Planck data in the determination of cosmological parameters. We consider the above mentioned multipole range in our analysis, while taking $f_{\mathrm{sky}} = 0.4$ as the sky fraction coverage. As for the noise modeling, we use the curves made available by the collaboration in \cite{SimonsObservatory:2018koc} for a baseline configuration (see their fig. 5) \footnote{\url{https://github.com/simonsobs/so_noise_models}}.

\item \textbf{LiteBIRD:} The LiteBIRD survey ~\cite{Matsumura:2013aja,Hazumi:2019lys,LiteBIRD:2022cnt} primarily constrains the large angular scales, providing sensitivity to the low multipoles of the CMB temperature and polarization spectra. In our case, we use Eq. \ref{eq:12} to model the instrument noise, with sky fraction coverage of $f_{\mathrm{sky}} = 0.7$, while $\sigma_{E,B}/\sqrt{2} = 4.17~\mu\mathrm{K}$-arcmin. The beam full width at half maximum is $\theta_{\mathrm{FWHM}} = 23$~arcmin, in agreement with the 140 GHz frequency channel as displayed in \cite{Paoletti:2019pdi}. In order to properly combine with Simons Observatory, we limit the multipole range as $2\leq\ell<40$, enough to cover the primordial part of the B-mode spectrum, as well as the reionization contribution to the E-mode CMB spectrum.

\item \textbf{Euclid:} The Euclid mission ~\cite{Euclid:2023pxu,euclid24}, launched in mid-2023, began collecting scientific data in early 2024, to provide three-dimensional clustering measurements described by the observed galaxy power spectrum. The survey aims to map a sky fraction of approximately $f_{\text{sky}} = 0.3636$, with its first public data release expected in October 2026~\cite{euclidqr25}. In this analysis, we focus exclusively on the galaxy clustering component, considering the spectroscopic part of the survey.

For the spectroscopic analysis, we consider scales up to $k_{\max} = 0.25\;h/\text{Mpc}$, corresponding to the conservative configuration defined in~\cite{euclid2025_sensi}, while taking $\sigma_z = 0.002$ for the spectroscopic redshift uncertainty. The likelihood implementation follows the official Euclid likelihood module available in \texttt{MontePython}~\cite{Euclid:2023pxu}.

\end{itemize}

\section{Results and Discussion}\label{Sec:4}

\begin{figure*}
    \centering
    \includegraphics[width=1.6\columnwidth]{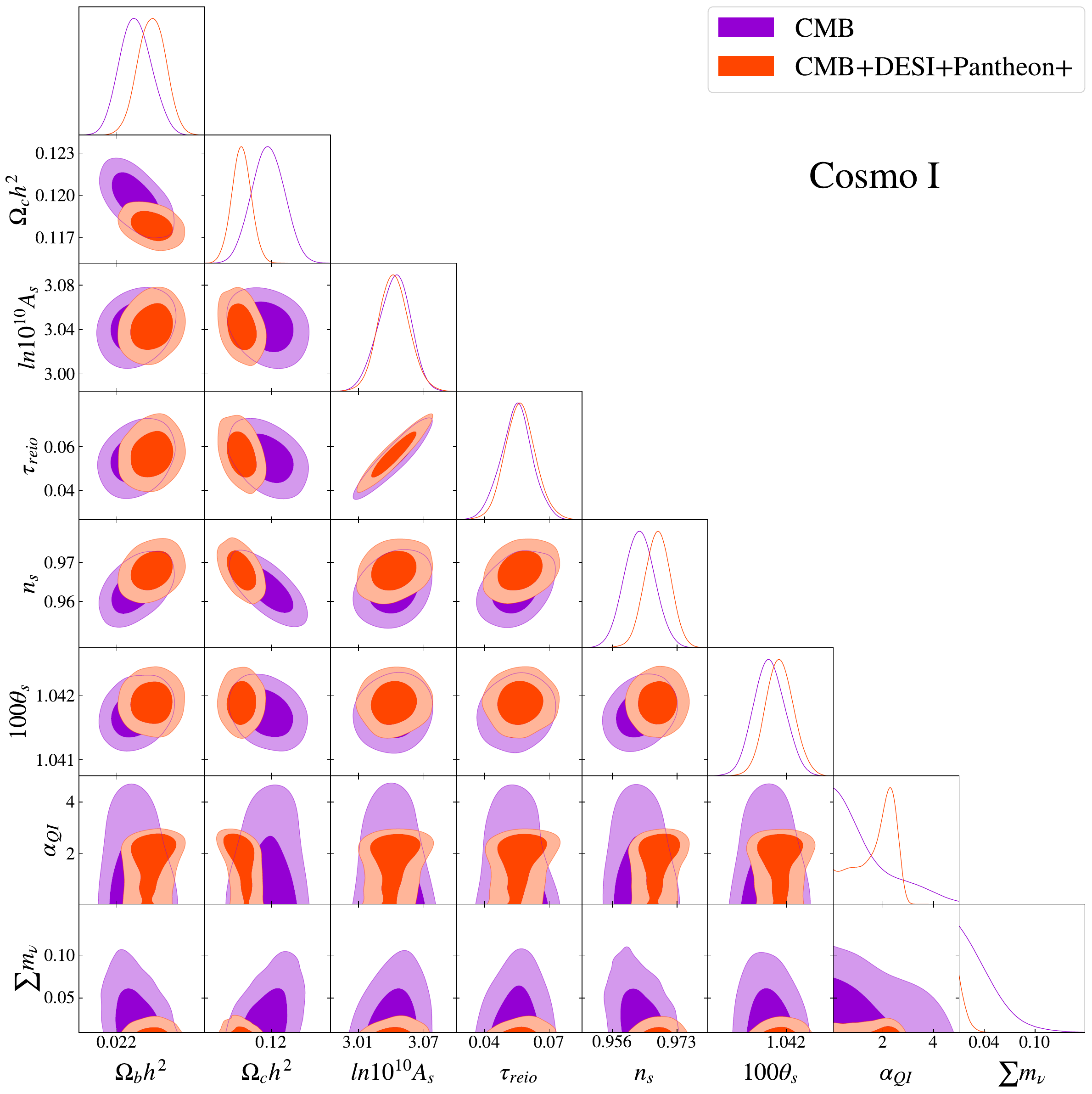}
    \caption{68\% and 95\% contours and normalized posteriors on the quintessential inflation model given by Eq.~\eqref{Eq:CosmI}. }
    \label{fig:1}
\end{figure*}

In Figure \eqref{fig:1}, we present the contour regions obtained for the $\mathrm{Cosmo\, I}$ model, and the corresponding $68\%$ constraints in Table~\eqref{Tab1}, in light of early- and late-time data. In particular, we explore two combinations of the datasets in our analysis. The first one considers the CMB temperature, polarization, and lensing data from Planck, while the second adds the BAO measurements from DESI's year two data release and the SNe Ia luminosity distances from Pantheon+ catalog.

\begin{table*}
\centering
\begin{tabular}{ l  c  c }

\hline \hline

 Parameter &    Simons Observatory + LiteBIRD  \,\,\,\,\,  &  Simons Observatory + LiteBIRD + Euclid \\
\hline
{$\Omega_\mathrm{b} h^2$}  & $0.000059$ & $0.0000503$ \\
{$\Omega_\mathrm{c} h^2$}  & $0.000819$ & $0.000407$ \\
{$100\theta_\mathrm{s}$}   & $0.000154$ & $0.000113$  \\
{$\tau_\mathrm{reio}$}  & $0.00200$ & $ 0.00194$ \\
{$n_\mathrm{s}   $}  & $0.00257$ & $0.00196$\\
{$\ln(10^{10} A_\mathrm{s})$}  & $0.00420$ & $0.00382$ \\
$\sum m_\nu $  & $0.0530$ & $0.0192$ \\
{$\alpha_\mathrm{QI}$}  & $0.479$ & $ 0.286$                         \\

\hline
\end{tabular}
\caption{1$\sigma$ forecasted constraints on the cosmological parameters.}
\label{Tab3}
\end{table*}

\begin{figure*}
    \centering
    \includegraphics[width=1.6\columnwidth]{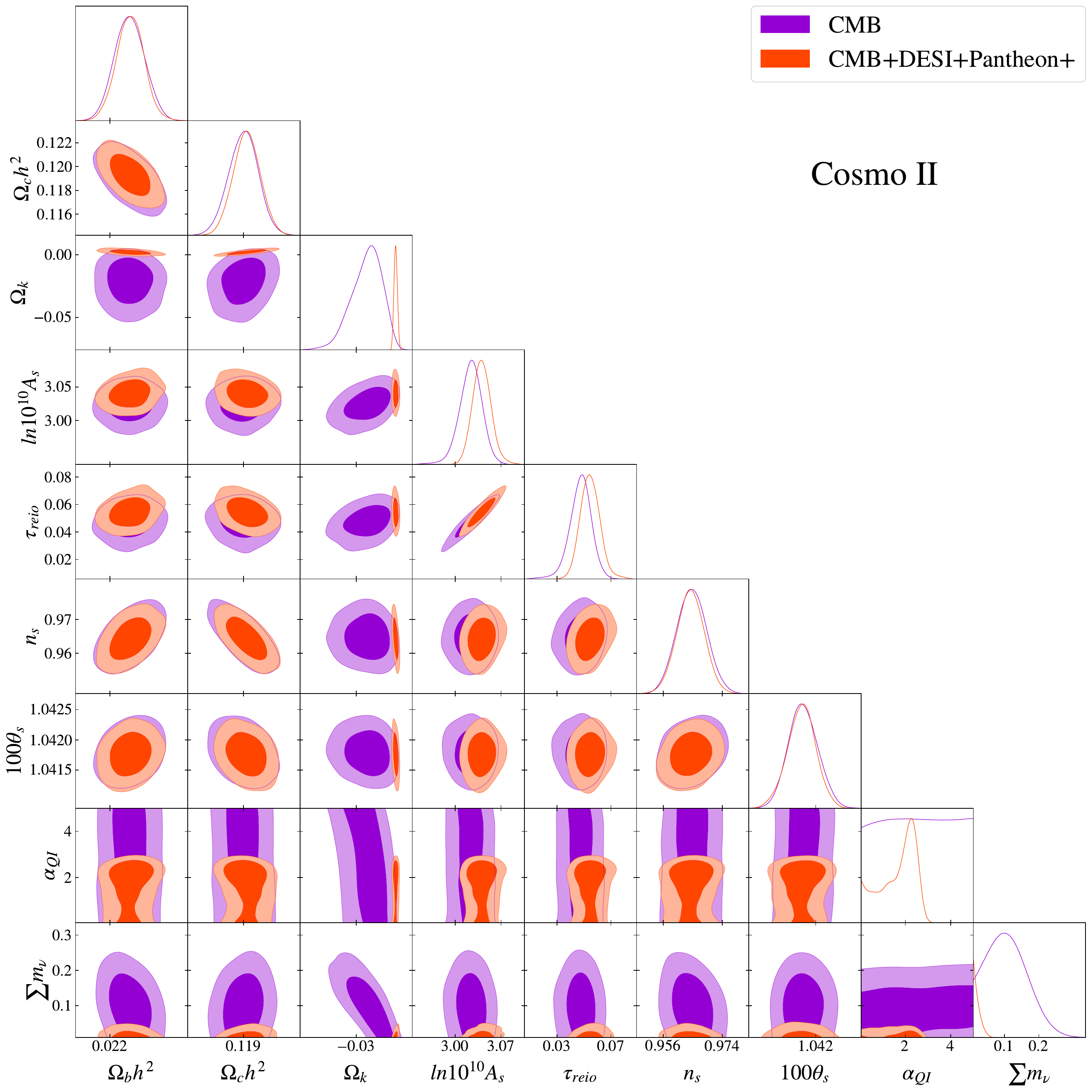}
    \caption{68\% and 95\% contours and normalized posteriors on the quintessential inflation model given by Eq.~\eqref{Eq:CosmII}. }
    \label{fig:2}
\end{figure*}

\begin{figure*}
    \centering
    \includegraphics[width=\columnwidth]{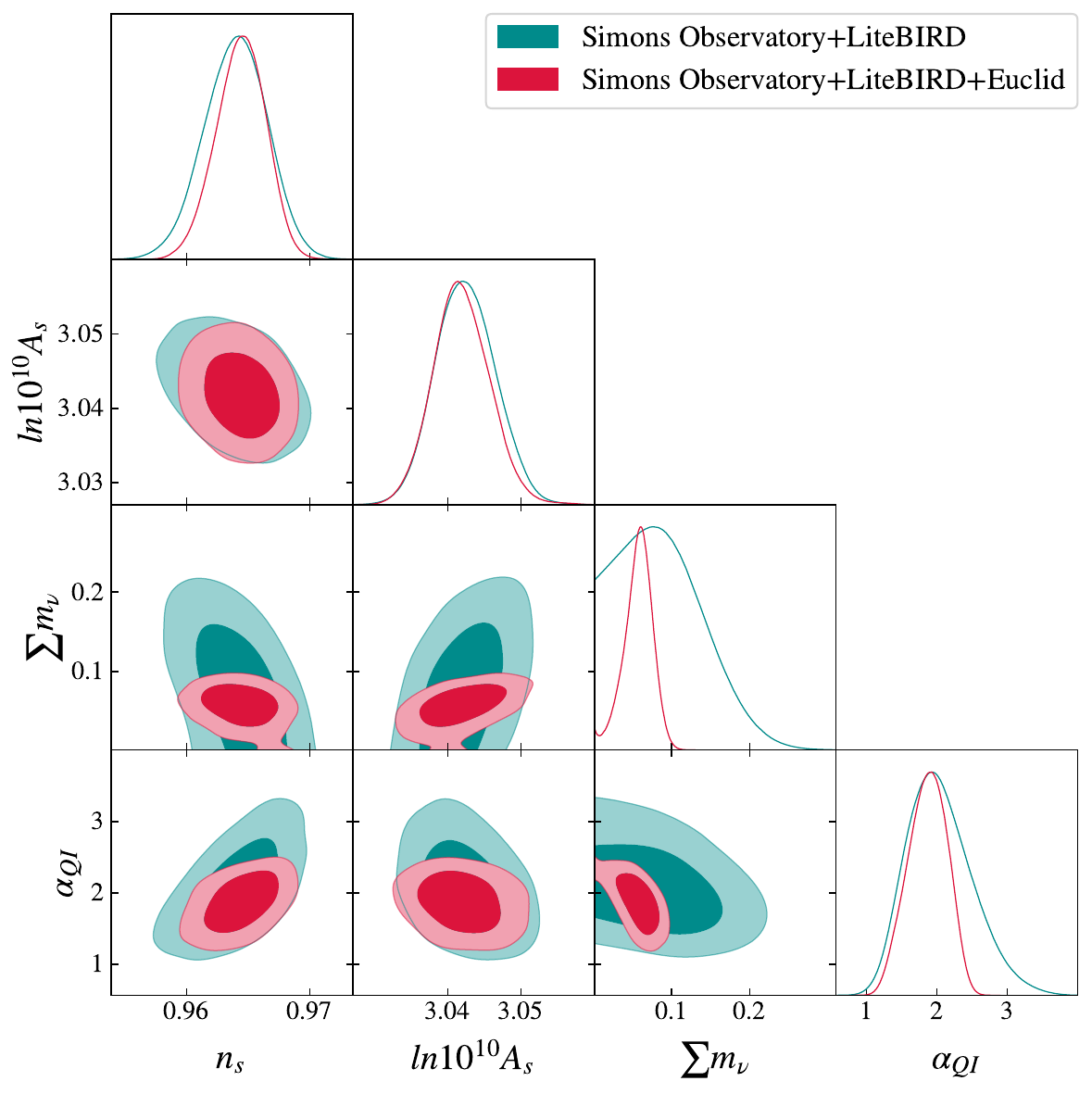}
    \includegraphics[width=\columnwidth]{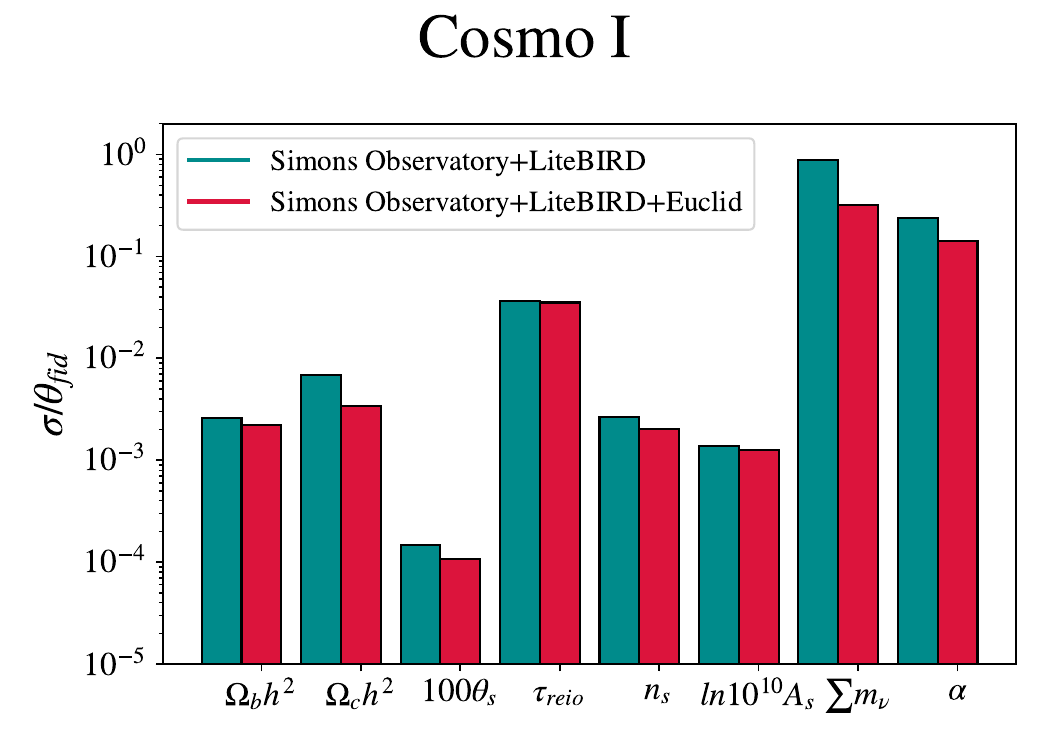}
    \caption{On the left panel, we show 68\% and 95\% contours and normalized posteriors on the forecasted sensitivity of the quintessential inflation `Cosmo I' model for the Simons Observatory, LiteBIRD and Euclid configurations, while the uncertainties are shown on the right panel. }
    \label{fig:3}
\end{figure*}

From the figure, one can observe a substantial improvement in parameter constraints when early- and late-time observables are analyzed jointly\textendash particularly for $\Omega_ch^2$, $\alpha_{QI}$, and $\sum m_\nu$. The CMB-only contours for $\sum m_\nu$ and $\Omega_ch^2$ reveal a positive correlation, reflecting the geometric degeneracy between $\sum m_\nu$, $\Omega_m$, and $H_0$~\cite{Loverde:2024nfi}. This degeneracy is partially broken when BAO distance measurements and supernova data are included in the analysis.

On the other hand, early-time observations alone are unable to constrain the curvature of the K\"ahler manifold, $\alpha_{QI}$. The CMB-only data imposes merely an upper bound, $\alpha_{QI} < 3.45$, at the $95\%$ C.L. This limitation is expected, since the constraining power of the CMB on this parameter originates from the absence of detected tensor modes, as described by Eq.~\eqref{eq:Slow_Rel}.

The situation changes significantly when late-time data are included. As illustrated in Figure 2 of \cite{Akrami:2020zxw} and discussed therein, the parameter $\alpha_{QI}$ is directly linked to the dark energy equation of state at low redshift. As a result, tighter constraints are obtained when information from the local Universe is considered. In particular, the joint analysis of CMB + DESI + Pantheon+ data yields $\alpha_{QI} = 1.70^{+1.00}_{-0.41}$, at $68\%$ C.L.\footnote{However, only an upper limit of $\alpha_{QI} < 2.64$ can be established at $95\%$ C.L.}.

The combination of early- and late-time cosmological probes also improves the constraints on the total neutrino mass, $\sum m_\nu$. In particular, the joint analysis of CMB, DESI, and Pantheon+ data yields an upper limit of $\sum m_\nu < 0.0670~\mathrm{eV}$, approximately $73\%$ tighter than the constraint obtained from CMB data alone, $\sum m_\nu < 0.252~\mathrm{eV}$. As shown in the figure, the marginal posterior distribution for $\sum m_\nu$ peaks near the lower prior boundary, indicating the presence of a prior-volume (or prior-weight) effect, as extensively discussed in \cite{DESI:2025ejh}. Since current cosmological observations still provide only an upper bound on $\sum m_\nu$, the posterior distribution remains partially sensitive to the adopted prior choice. In our analysis, we restrict the parameter space to physically motivated positive values of $\sum m_\nu$ and tested the robustness of the inferred bounds by varying the upper limit of the prior range. We find that the resulting constraints remain qualitatively stable under these variations, although some residual prior dependence is expected given the absence of a significant detection of the neutrino mass scale.

Furthermore, the total neutrino mass affects the late-time cosmic expansion, modifying the transverse comoving distance to the last scattering surface and altering the angular scale of the acoustic peaks. Notably, the tension between the distance scales inferred from CMB and BAO observations can be alleviated by reducing $\sum m_\nu$ within the $\Lambda$CDM framework~\cite{DESI:2025ejh}. This mechanism contributes to the tighter constraints obtained from the combined dataset.
Previous analyses involving dynamical dark energy parameterizations suggest that such bounds can be relaxed in those models \cite{Rodrigues:2025hso}, due to degeneracies between $\sum m_\nu$ and the dark energy equation of state, $\omega_{\phi}(z)$ \cite{Hannestad:2005gj, RoyChoudhury:2019hls}. This degeneracy is not present in the quintessential inflation model, where the unified description of early- and late-time expansion naturally breaks the $\sum m_\nu$–$\omega_{\phi}$ correlation, enabling bounds that are competitive with those from the most restrictive scenarios.

In order to further explore the effects of possible degeneracies, we also perform the statistical analysis of the $\mathrm{Cosmo\, II}$ model, which considers the role of a non-zero space-time curvature in the background geometry. We present our results in Figure~\eqref{fig:2} and Table~\eqref{Tab2}. The figure reveals a strong anti-correlation between $\Omega_K$ and $\sum m_\nu$ in the CMB-only contour. This degeneracy is sourced from the angular distance to the last scattering surface~\cite{Howlett_2012}, being partially broken by the late-time data. The constraints are significantly improved when the BAO and supernovae data are included in the analysis. In particular, an improvement of $81.6\%$ is obtained in the constraint of the total neutrino masses, and the $1\sigma$ interval is settled for the quintessential parameter, $\alpha_{QI}=1.69^{+1.0}_{-0.44}$. 

We agree that the result is interesting, but the tension between the constraints from CMB-only and CMB+BAO is not very large when considering the $1\sigma$ uncertainties (approximately $1.6\sigma$). The same qualitative behavior is observed in the $\Lambda$CDM$+\Omega_K$ model, as reported by the Planck collaboration~\cite{Planck:2018vyg}, with $\Omega_K = -0.0106 \pm 0.0065$ from CMB-only data and  $\Omega_K = 0.0007 \pm  0.0019$ from CMB+BAO. Our results for $\Omega_K$ are also consistent with those obtained by the DESI collaboration~\cite{DESI:2025zgx}, where $\Omega_K = 0.0023 \pm  0.0011$ was reported for the $\Lambda$CDM$+\Omega_K$ analysis using CMB+BAO data. As discussed in~\cite{Chen:2025mlf} and \cite{DESI:2025ejh}, the comoving distance to the last scattering surface depends on the curvature of the Universe. The MCMC analysis for the CMB+BAO data seems to prefer a shorter distance.

It is also possible to note, from Table~\eqref{Tab2}, that late-time data displaces the constraints on the curvature to the positive range at $1\sigma$, while still encompassing the zero point in the 95$\%$ confidence level interval, $\Omega_K=0.0021^{+0.0027}_{-0.0025}$. The same qualitative behavior is observed in the $\Lambda$CDM$+\Omega_K$ model, as reported by the Planck collaboration~\cite{Planck:2018vyg}. Our results for $\Omega_K$ are also consistent with those obtained by the DESI collaboration~\cite{DESI:2025zgx}, which reported $\Omega_K = 0.0023 \pm 0.0011$ in the $\Lambda$CDM$+\Omega_K$ analysis using CMB+BAO data. 
This shift can be understood as a change in the comoving distance to the last scattering surface. DESI data seem to prefer a shorter distance due to its preference for a dynamical dark energy. Such an effect is degenerated with the curvature of the Universe. Therefore, the MCMC analysis of the CMB+BAO data finds a new equilibrium point that balances the relevant parameters.

Furthermore, the degeneracy in the background evolution of  $\Omega_K$ and $\sum m_\nu$ is not entirely broken by late-time data, which results in a loosening of the upper limit for the total neutrino masses with respect to what is obtained for the $\mathrm{Cosmo\, I}$ model. As a result, both the normal and inverted hierarchies of the neutrino masses are possible in the $\mathrm{Cosmo\, II}$ scenario. A similar result is reported in Ref.~\cite{Chen:2025mlf}, using CMB + DESI BAO data. Allowing the spatial curvature to deviate from zero significantly relaxes the constraints on $\sum m_\nu$ within the $\Lambda$CDM model, while revealing a preference for negative $\Omega_K$. Only in the analysis of the $\mathrm{Cosmo\, II}$ model with exclusively CMB data did the posterior of $\sum m_\nu$ peak in the positive parameter region, while the inclusion of the BAO data in the analysis seems to shift the posterior to the negative range\footnote{Although we do not consider a negative prior for the total neutrino mass, we consider the peak of the posterior at 0 to indicate a preference of the models for negative values of this parameter.}. One may interpret this result as symptomatic of the discrepancy between the distance scales obtained from the DESI-BAO measurements and those from the CMB data, as discussed in \cite{DESI:2025ejh}.

Finally, we test the models against the configurations of future cosmological surveys. Once again, we perform the MCMC analysis for simulated data with the settings of the Simons Observatory and LiteBIRD experiments (early-time data, CMB) and a joint analysis considering the configurations of the Euclid satellite (late-time data, galaxy survey). In Fig.~\ref{fig:3}, we present the 2D contour regions for a representative subset of parameters of the Cosmo I model (left panel) and the corresponding improvement in their uncertainties (right panel), represented by the ratio of the uncertainties shown in Table~\ref{Tab3} with the fiducial parameters chosen for the analysis, being $(\Omega_b h^2,\Omega_ch^2,100\theta_s,\tau_{reio},n_s,\ln(10^{10}A_s),\sum m_\nu,\alpha)=(0.02277,0.12,1.04110,0.055,0.965,3.045,0.06,2)$.

At first glance, one may note the improvement in precision when considering early- and late-time simulated data jointly, especially for $\alpha_{QI}$ and $\sum m_\nu$ parameters. Not surprisingly, the precision has also increased in relation to the previous MCMC analysis using current data. In particular, for the $\alpha$-attractor parameter, $\alpha_{\mathrm{QI}}$, we obtain the forecasted $1\sigma$ uncertainty of $ 0.286 $ (Simons + LiteBIRD + Euclid), which represents an improvement in precision of approximately $72\%$  with respect to the most constraining current case (real data), i.e, the flat model with CMB + DESI + Pantheon+. For the total neutrino mass, we also find a significant increase in sensitivity, with the expected error of $0.0192$ for the combined configurations of the future galaxy survey and CMB observations, corresponding to an improvement of roughly $8.6\%$ from the constraining power obtained with the analysis using real data\footnote{The basis for this comparison is the $1\sigma$ uncertainty obtained from analysis of the Cosmo I model with CMB + DESI + Pantheon+ data, which has returned an integrated $1\sigma$ posterior uncertainty of $0.021$ for the total mass of neutrinos.}. We also note that our results for $\sum m_\nu$ are competitive with those obtained in similar analyses~\cite{Rodrigues:2025zvq,Qin:2025nkk,Chudaykin:2019ock}. In particular, using a Fisher matrix approach, Ref.~\cite{Rodrigues:2025zvq} reported a $2\sigma$ sensitivity of $0.033$ eV for the total neutrino mass within the $\Lambda$CDM $+\sum m_\nu$ framework, considering the J-PAS galaxy survey configuration combined with CMB and SNe Ia observations. This constraint is approximately $15\%$ tighter than ours.

\section{Conclusions}\label{Sec:5}

The ever-increasing volume of high-quality data has enabled cosmological surveys to probe both the early- and late-time Universe with unprecedented precision, placing stringent constraints on some of the most fundamental theories in physics. Examples include the inflationary paradigm and neutrino physics. In the latter case, cosmological upper bounds on the sum of neutrino masses have progressively approached the lower limits inferred from neutrino oscillation experiments. Although no significant tension is currently observed among existing data sets, forthcoming survey configurations are expected to substantially improve these constraints, potentially determining the mass scale of active neutrinos or ruling out some of the standard scenarios within the cosmological model.

In this work, we tested the total neutrino mass in the framework of a quintessential inflationary model based on the $\alpha-$attractor scenario. Data from early- and late-time Universe were considered in order to break some degeneracies in the model parameters. We focus on two configurations of the cosmological model. The best constraint for the neutrino masses is obtained for the $\mathrm{Cosmo\, I}$ model, $\sum m_\nu < 0.067$ eV, which seems to exclude the inverted ordering while still being in agreement with the normal one. We emphasize that, with respect to dynamical dark energy models, this is one of the most restrictive results for total neutrino masses in the current literature\footnote{For comparison, this restriction is $6\%$ more restrictive than that obtained for the thawing parameterizations in \cite{Rodrigues:2025hso} and $48\%$ more restrictive than that obtained for the CPL parameterization from DESI collaboration~\cite{DESI:2025ejh}.}. On the other hand, the limits for the $\mathrm{Cosmo\, II}$ model are consistent with both orderings, $\sum m_\nu < 0.116$ eV. Also, $68\%$ C.L. constraints for $\alpha$ were obtained on both models (see \eqref{Tab1} and \eqref{Tab2}), while only upper limits were obtained at $95\%$ C.L. In what concerns curvature, the CMB measurements returned negative values for $\Omega_K$, with the late-time data displacing this parameter into the positive range. Although the $95\%$ C.L. constraints on both models are still consistent with the flat Universe, which is an important consistency test for inflation itself, the apparent tension between these constraints may point to an underlying discrepancy between the scales measured from early and late Universe.

We also test the model against the configurations of future cosmological surveys. The results project a notable improvement in parameter sensitivity. For the $\alpha$-attractor parameter, we obtain an improvement in precision of approximately $72\%$ with respect to the most constraining current case. For the total neutrino mass, we also find a significant increase in sensitivity, corresponding to an improvement of roughly $8.6\%$ in constraining power. We suppose that this modest improvement in the constraints on the total neutrino masses is mainly due to the already exceptional precision of the Planck data, particularly the lensing potential reconstruction. Nevertheless, it is worth noting that this increase in precision may help determine the scale of neutrino masses and, ultimately, establish the neutrino hierarchy.

Although current experiments have not yet provided a clear determination of the origin of neutrino masses, future cosmological surveys hold significant potential to address long-standing questions regarding the fundamental nature of these particles. Meanwhile, forecast constraints on the sum of neutrino masses suggest that this parameter may act as a powerful probe of viable quintessence scenarios.

\section*{Acknowledgements}
JR thanks the Funda\c{c}\~ao de Amparo \`a Pesquisa do Estado do Rio de Janeiro (FAPERJ) grant No. E-26/200.513/2025 and Conselho Nacional de Desenvolvimento Científico e Tecnológico (CNPq) grant No. 406718/2025-3. GR is supported by the Coordenação de Aperfeiçoamento de Pessoal de Nível Superior (CAPES). FBMS is supported by Conselho Nacional de Desenvolvimento Científico e Tecnológico (CNPq) grant No. 151554/2024-2. S.S.C. acknowledges support from the Istituto Nazionale di Fisica Nucleare (INFN) through the Commissione Scientifica Nazionale 4 (CSN4) Iniziativa Specifica ``Quantum Fields in Gravity, Cosmology and Black Holes'' (FLAG), from the Fondazione Cassa di Risparmio di Trento e Rovereto (CARITRO Foundation) through a Caritro Fellowship (project ``Inflation and dark sector physics in light of next-generation cosmological surveys''), and CNPq Grant No. 446810/2024-0. JSA is supported by CNPq Grant No. 307683/2022-2 and by FAPERJ Grant No. 299312 (2023). We also acknowledge the use of the \texttt{Cobaya}, \texttt{CLASS}, and \texttt{GetDist} codes. This work made use of the National Observatory Data Center (CPDON).

\section*{Data availability}
The modified version of the \texttt{CLASS} code we used to perform our analyses will be made available upon request.

\bibliography{references}

\label{lastpage}

\end{document}